\documentclass[11pt,twoside]{article}

\usepackage{asp2006}
\usepackage{epsf}

\begin{document}
\title{Accretion Disks in AGNs}   
\author{Omer Blaes}   
\affil{Physics Department, University of California, Santa Barbara, CA 93106, USA}    

\begin{abstract} 
I briefly review the theoretical models of radiatively efficient, geometrically
thin and optically thick accretion disk spectra that currently exist for AGN.
I then discuss three recent observational developments that have real potential
to teach us about the physics of these flows.  Finally, I present results
on the most recent, thermodynamically consistent simulations of
magnetorotational turbulence and discuss what these simulations are suggesting
about the vertical structure of accretion disks.
\end{abstract}


\section{Current Standard Accretion Disk Models of AGN}   

Accretion disks around supermassive black holes continue to be the generally
accepted paradigm for the central engine of AGN.  I focus in this review on
``standard disks'': those that are geometrically thin and optically thick.
These models are,
by construction, radiatively efficient.  For radiatively inefficient disks,
see the papers by Mineshige and by Yuan in these proceedings.
Currently all theoretical models of the photon emission of accretion disks
(radiatively efficient or otherwise) are based on
the alpha prescription for the anomalous stress introduced by \citet{omb_sha73}.
A large number of such models for standard disks have
been constructed over the years,
e.g. \citet{omb_mal83,omb_sun89,omb_lao89,omb_lao90}, to name just a few.

The most sophisticated of these models have been developed in a
series of papers by \citet{omb_hub97,omb_hub98} and
\citet{omb_hub00,omb_hub01}, and are available
to the community through the IDL package AGNSPEC, written by Ivan
Hubeny.
The models are constructed by first
dividing the accretion disk into a set of discrete annuli, and then computing
a one-dimensional vertical interior and atmosphere structure based on
certain assumptions in each annulus.   Using the locally calculated
spectrum, photon geodesics from the disk to
infinity are computed to then produce a predicted model spectrum for an
observer at some viewing angle.  Some aspects of the physics
of accretion disks are treated very well by the AGNSPEC models.  They
fully account for the relativistic radial structure of the disk
\citep{omb_nov73,omb_rif95} and
relativistic Doppler shifts, gravitational redshifts and light bending in
a Kerr spacetime \citep{omb_ago97}.
The models include a detailed non-LTE treatment of all
abundant elements and ions in the disk, and include continuum opacities due
to bound-free and free-free transitions of these ions, as well as
Comptonization.  (No lines are included in the models, however.)

There are many ad hoc assumptions that continue to plague these models,
however.  First there are assumptions that relate to the radial structure of
the disk.  The disk is assumed to be stationary, with a zero torque inner
boundary condition at the innermost stable circular orbit.  Recent global
simulations of MRI turbulent flows around black holes suggest instead that
substantial torques on the disk can be exerted by material in the plunging
region through magnetic field lines that connect this material to the disk
(e.g. \citealt{omb_kro05}).
The models also adopt the \citet{omb_sha73} alpha prescription
for the anomalous stress, with the stress being proportional to the total
thermal (gas plus radiation) pressure and the alpha parameter assumed constant
with radius.  This is done even though such models, if taken literally,
are thermally and ``viscously'' unstable when radiation pressure dominates
gas pressure \citep{omb_sha76}.  Reprocessing of radiation that leaves the disk
at one point and returns to a different point due, e.g. to light bending, is
neglected.  This can be an important effect particularly for rapidly spinning
black holes \citep{omb_cun76}.

There are perhaps even more serious assumptions regarding the vertical
structure at any radius.  This vertical structure is first assumed to depend
only on height, i.e. it is entirely one dimensional.  As
I discuss below, this is probably not accurate as substantial inhomogeneities
probably exist. The structure is also assumed to be symmetric
about the equatorial plane of the black hole: no warps or bending waves are
considered, nor are time-dependent asymmetries induced by the turbulence
accounted for \citep{omb_hir06}.  The disk is assumed to be in vertical
hydrostatic
equilbrium with gas and radiation pressure being the only means of support
against the vertical tidal field of the black hole.  Magnetic forces are
neglected.  Just as a model of a
stellar interior requires
specification of the sources of nuclear energy, a model of the vertical
structure of the disk requires specification of the vertical profile of
turbulent dissipation.  The AGNSPEC models assume that the dissipation per
unit volume tracks the density, i.e. the dissipation per unit mass is
constant.  Heat is also assumed to be transported vertically through radiative
diffusion.  These assumptions are not consistent in a radiation pressure
dominated annulus:  constant dissipation
per unit mass plus radiative diffusion immediately leads to a density which
is constant with height if the opacity is dominated by electron scattering (a
good approximation).  This in turn is convectively unstable \citep{omb_bis77}.

It is probably for many of these reasons that the models do not do a very good
job in reproducing some aspects of the observations (e.g. \citealt{omb_kor99}).
For example, they do not produce X-rays.  They produce a break at the Lyman
limit of hydrogen but not quite like what is observed \citep{omb_bla01}, nor
do the expected trends in far ultraviolet continuum slopes agree with
observation \citep{omb_sha05}.  The models
cannot explain the observed near simultaneous optical/ultraviolet variability
\citep{omb_all85,omb_cou91} because they do not include reprocessing.

In spite of the fact that the physics of these models is so uncertain, I should
emphasize that exactly the same type of modeling applied to {\it stellar} mass
black hole accretion disks (BHSPEC, \citealt{omb_dav06}) results in very
good fits to the spectra of black hole X-ray binaries in the high-soft/thermal
dominant state \citep{omb_ddb06}.  For LMC~X-3, for example, where we know the
distance and the black hole mass, we can fit the accretion disk spectrum
very well over a range of luminosities with fixed spin and inclination.  The
models only fail at the very highest luminosities (Eddington ratio greater
than of order 0.7), where the assumption of a geometrically thin disk is
almost certainly wrong.  Either the ad hoc physical assumptions that go into
the disk models are not too wrong, or the spectra are robust to
changing these assumptions.

\section{Some Recent Observational Developments that Provide New Constraints
on Disk Physics}

I would now like to mention three recent observational developments that are
exciting because they provide us with new empirical constraints on what
is going on in AGN accretion disks.

\subsection{Spectropolarimetry}

For good reason, the broad and narrow emission line regions in
AGN have attracted enormous attention from the community for years.  However,
to a parochial theorist like myself who is primarily interested in the accretion
flow very near the black hole, emission from these regions merely contaminates
our view of the innermost parts of the central engine.  In particular, the
numerous broad and narrow emission lines, the small blue bump, and the
infrared thermal emission from dust make an accurate determination of the
shape of the big blue bump (which is supposed to originate in the accretion
disk) very difficult.

Fortunately, all this contaminating emission appears to be unpolarized in a
subset of quasars, whereas the underlying big blue bump is polarized.  The
reasons for this are unclear, but it appears that there is an electron
scattering medium
enveloping the central accretion flow but lying entirely within the broad
line region in these quasars.  If one measures the polarized flux in these
quasars, all the contaminating emission is removed and the shape of the
underlying continuum is revealed.  In particular, an absorption feature which
is almost certainly a Balmer edge lies underneath the small blue
bump \citep{omb_kis03,omb_kis04}.  This is the first direct evidence that
the big blue bump originates from thermal emission from an optically thick
medium!  This technique has also been used to measure the long wavelength,
near-infrared shape of the big blue bump underneath the thermal emission
from dust \citep{omb_kis05}.  Kishimoto discusses these results in much more
detail in his paper in these proceedings.

\subsection{Microlensing}

While it might be possible to directly image the very central part of the
accretion flow around the black holes in the Galactic center and M87
in the not so distant future with VLBI \citep{omb_fal00,omb_bir02},
the angular sizes of
the radiatively efficient black hole accretion disks that we know about in
both AGN and X-ray binaries are hopelessly small for imaging to really be
practical.  Fortunately, nature has provided us with a cosmic magnifying
glass in the form of gravitational lensing.  Many examples of multiply imaged
quasars are now known and are fit reasonably well by models of the foreground
lensing galaxy (the macrolens).  Depending on the source size at a particular
observed wavelength, each of the quasar images can in turn be microlensed
by stars that reside in the foreground macrolens galaxy.  If the source (the
quasar) has a size scale much less than the Einstein radius of a typical
stellar microlens, then the flux ratios of the different quasar images will
typically differ markedly from the predictions of the macrolens model.  They
will also vary with time in a manner that has nothing to do with the intrinsic
variability of the quasar.  On
the other hand, if the source is extended compared to the microlens Einstein
radius, there will be little differential magnification and the flux ratios
should agree with the macrolens model.  Microlensing therefore probes the
size of quasar emission regions on scales comparable to the Einstein radius
of a stellar microlens.  For cosmologically distant lenses and sources, this
corresponds to angular resolutions of several microarcseconds, or linear
scales in the quasar $\sim5\times10^{16}$~cm, for a solar mass microlens.

\citet{omb_rau91} were the first to use this technique. For the Einstein Cross,
the quadruply lensed quasar where microlensing was first discovered
\citep{omb_irw89}, they
found that the optical (really near ultraviolet) emission regions were
{\it smaller} than expected
for thermal emission from an accretion disk.
More modern work in this system and other
gravitational lenses has generally reached the opposite conclusion:  the
near ultraviolet emission regions are {\it larger} than predicted by simple
disk models.  In particular, \citet{omb_poo06} have found that near
ultraviolet half-light radii predicted by multitemperature blackbody disks
are a factor 3 to 30 times too small than what is necessary to explain the
observed microlensing flux ratios in ten quasars, including the Einstein Cross.
Using a different analysis of the microlensing statistics and (hopefully
better) black hole masses based on emission line widths,
\citet{omb_koc06} and \citet{omb_dai06} also find that the near ultraviolet
emission regions are larger than would be expected for the size of the
annulus with the corresponding effective temperature.  However, they also find
that the size of the emission region has roughly the right scaling with black
hole mass between different sources.

It is not yet clear what causes the discrepancies \citep{omb_dai06}.
Contamination
of the observed fluxes by the small blue bump is clearly a worry, as this
originates from larger scales than the disk.  Reprocessing of inner disk
radiation will also increase the size of the near ultraviolet emitting annuli.
Disk atmosphere models such as AGNSPEC show significant deviations from
local blackbody emission:  the theoretical spectrum is harder, which would
make the size of the emission region at a given wavelength larger than expected
from a local blackbody disk.  Whatever the cause, this observation technique
provides a unique probe of the size scales of the emission regions, and I
strongly encourage observers to extend it across multiple wavelengths.

\subsection{SDSS Spectra with Black Hole Masses}

As discussed by Kaspi and Peterson in these proceedings, reverberation
mapping leveraged by the broad line region radius/continuuum luminosity
correlations has given us a simple method to measure approximate black
hole masses, and this can be done for huge numbers of quasars in the
Sloan Digital Sky Survey (SDSS) sample.  As a result, one can now look at
continuum colors as a function of luminosity $L$ and black hole mass $M$
and see whether the variations match the trends predicted by accretion disk
theory.

\citet{omb_bon06} have taken a first stab at this by looking at
colors at three pairs of rest frame wavelengths (5100\AA/4000\AA,
4000\AA/2200\AA, and 2200\AA/1350\AA) versus a measure of the expected
maximum disk effective temperature $T_{\rm max}\propto L^{1/4}M^{-1/2}$.
They found that the 5100\AA/4000\AA\/ colors predicted by the AGNSPEC models
agreed with the SDSS data at all $T_{\rm max}$.  The 4000\AA/2200\AA\/ colors
agree at low values of $T_{\rm max}$, but not at high, and the 2200\AA/1350\AA\/
colors agree only at the lowest values of $T_{\rm max}$.  The sense of
the disagreement in the discrepant regions is that the data is redder than
the AGNSPEC predictions.  Internal reddening by dust in the quasars is
therefore a worry, but
they argue that this cannot explain the observed discrepancies.
Interestingly, they find that the value of $T_{\rm max}$ which marks the
point where the observed 4000\AA/2200\AA\/ colors depart from the model
predictions happens to correspond
to an Eddington ratio of 0.3, beyond which the standard geometrically thin
disk assumption may break down.  ``Slim'' disk physics may
therefore be playing a role (see the paper by Mineshige in these proceedings).

As \citet{omb_bon06} acknowledge in their paper, the combination of luminosity
and mass that appears in $T_{\rm max}$ actually depends empirically only on
the measured width of the Mg~II line used in their analysis, and not on the
measured luminosity.  It would be nice to extend their work to a truly
two-dimensional analysis with respect to luminosity and black hole mass.

\section{Some Recent Theoretical Developments:  $\alpha$ Begone!}

The \citet{omb_sha73} alpha prescription that has plagued accretion disk models
for decades is slowly being banished to oblivion.  Thanks to the discovery
of the magnetorotational instability (MRI, \citealt{omb_bal91}) and the ever
increasing
power of numerical simulations, the theoretical community is starting to
rebuild accretion disk theory from scratch.  One of the most exciting
developments is that radiation magnetohydrodynamic simulations are telling
us how the vertical structure might really work.  Such simulations are possible
in a local vertical slice through a disk, including the vertical gravity
and the MRI turbulence, capturing the dissipation of the turbulence as heat,
and allowing radiation to diffuse out of the box so that cooling can occur.
Both the dynamics and the thermodynamics of the turbulence are therefore
captured in the simulations.

Three such simulations have been completed so far.  The first was by
\citet{omb_tur04}, who modelled a radiation pressure dominated annulus and found
that it reached a thermal equilibrium, albeit with large fluctuations.  There
was no evidence of a thermal instability, but there was significant vertical
expansion of the disk during a heating phase which caused approximately
half the mass of the annulus to leave the grid.  The dissipation per unit
volume inside the annulus did not track the density, and the average
vertical structure was convectively stable.  Time step constraints on the
radiation diffusion solver required the use of a rather high density
floor within the simulation which meant that every zone was optically thick:
there was no photosphere within the simulation domain.  The results of this
pioneering simulation are tantalizing, but they are not yet definitive.

A number of improvements have since been made to the code used by
\citet{omb_tur04} so that it conserves energy much better and also runs
more efficiently and faster, allowing the density floor to be reduced to
a level where the photosphere can now be included within the simulation
domain.  A truly radiation pressure dominated annulus has not been
successfully run yet with the new code, however, possibly because the
standard \citet{omb_sha73} vertical structure assumptions are so far from
reality that using them as an initial condition leads to very large
transients.  A simulation of a gas pressure dominated annulus was published
by \citet{omb_hir06}.  A new simulation with comparable gas and radiation
pressures at the midplane has now been completed \citep{omb_bla06b,omb_kro06}.
It is probably the best simulation we have so far
of the conditions that might exist in black hole accretion disks, although
it is still not radiation dominated.

Figure 1 shows the time variation of the energy content in the box for this
new simulation \citep{omb_kro06}.  The average cooling time in the annulus
is only $\sim7$ orbits, and clearly the energy content of the box goes through
very large fluctuations on time scales much longer than this.  The nominal
\citet{omb_sha76} thermal instability criterion that total radiation pressure
exceed three fifths of the total thermal pressure is violated around 90
orbits and also (slightly) around 300 orbits.  Yet there is no evidence of
a thermal runaway.  The reason for this appears to be that the
volume-integrated dissipation rate is approximately proportional to the
square root of the total energy content during the hottest epochs, while
the cooling rate is proportional to the total energy content.  These
scalings therefore lead to thermal stability.  It is far from clear that
this will extend to regimes of much higher radiation pressure, however.

\begin{figure}[!ht]
\plotone{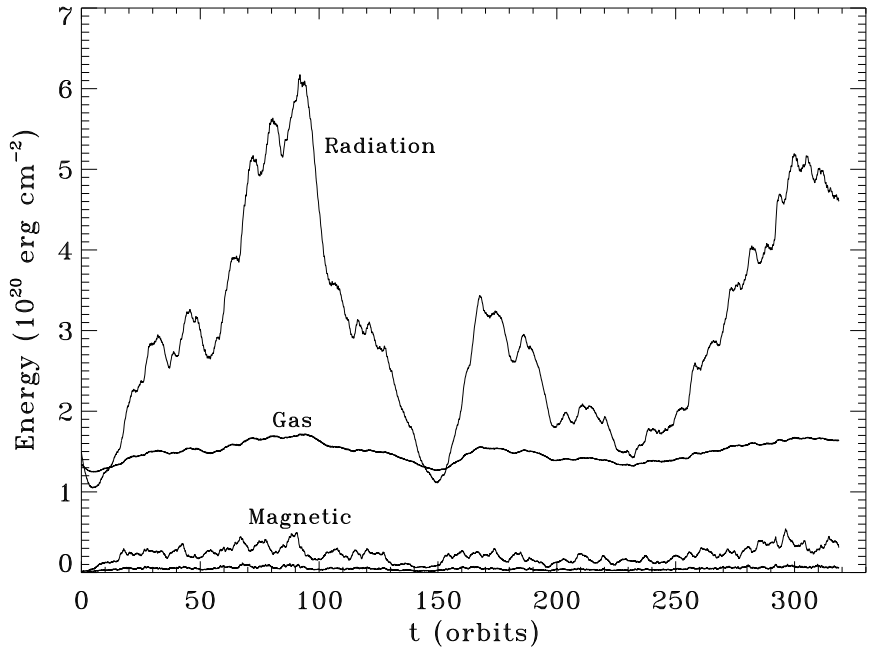}
\caption{Time variation of the total energy content for the new simulation
with comparable gas and radiation pressures.  From top to bottom, the curves
represent the vertically integrated radiation energy, gas internal energy,
magnetic energy, and fluid kinetic energy, respectively.}
\end{figure}

That describes the thermal equilibrium, but the hydrostatic equilibrium is
also interesting.  Away from the midplane, vertical magnetic forces dominate
gas and radiation pressure gradients in supporting the annulus against
gravity.  This was also seen in the gas pressure dominated simulation
by \citet{omb_hir06}.  This has immediate spectral implications as it
produces a larger density scale height compared to what would exist if
only gas and radiation pressure gradients were taken into account.  This
results in a lower density at the effective photosphere, producing increased
ionization and reduced absorption opacity.  This generically results in
a harder emergent spectrum \citep{omb_bla06a}.

However, one-dimensional stellar atmosphere modeling does not do the simulation
justice \citep{omb_bla06b}.  As shown in Figure 2, strong horizontal density
fluctuations (up to a factor 100) exist as deep as the effective photosphere
of the horizontally averaged structure and even deeper.
Some of the high density regions are reminiscent of the shock train structure
predicted in the nonlinear development of the photon bubble instability
\citep{omb_beg01}, but the expected correlations between radiation flux
and fluid velocity do not appear to match this instability.  On the other
hand, the transverse Parker instability is clearly present: high (low) density
regions at depth are strongly correlated with upward (downward) magnetic
tension forces.

At least to the extent that thermalization of photons with matter occurs
through true absorption and emission processes (as opposed, say, to Compton
scattering) the large density fluctuations will tend to enhance the
thermalization of photons, leading to a more blackbody-like, softer emergent
spectrum \citep{omb_dav04,omb_beg06}.

\begin{figure}[!ht]
\plotone{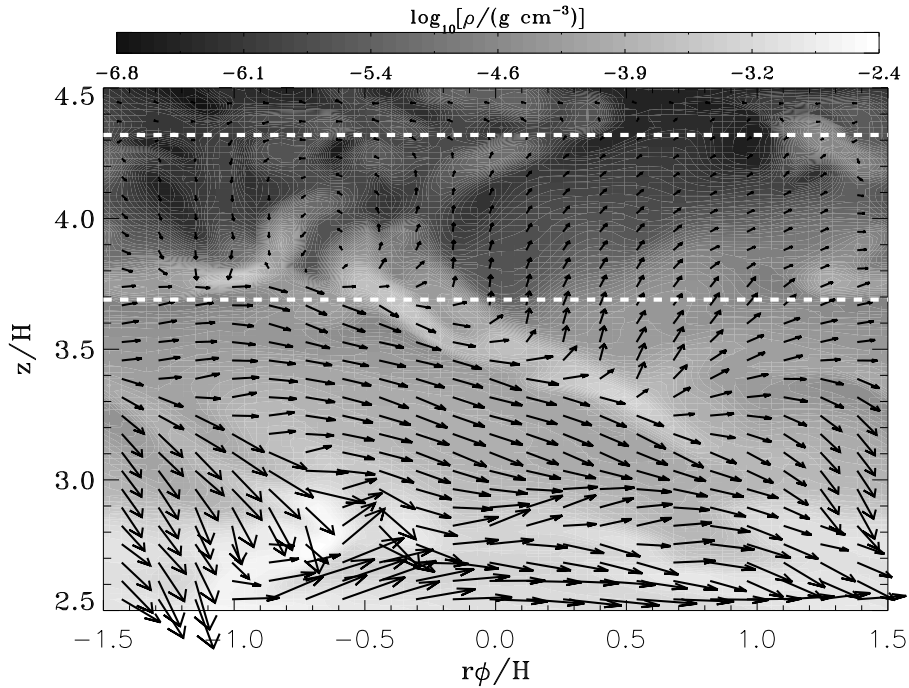}
\caption{Density (gray scale) and projected magnetic field (arrows) in a
constant radial slice through the subphotospheric upper layers of the
simulation at 90 orbits.  The upper and lower horizontal dashed lines show
the Rosseland mean and effective (i.e. thermalization) photospheres,
respectively, of the horizontally averaged structure.}
\end{figure}

The complex photosphere will also presumably reduce the degree of polarization
of the emergent radiation, as the atmosphere no longer has the plane-parallel
symmetry used in simple models \citep{omb_cha60}.  This has of course been
a long-standing problem with disk models:  electron scattering dominated slabs
generally produce too high a polarization which is also at the wrong position
angle \citep{omb_sto79,omb_ant88}.  The inhomogeneities that we see in the
simulation resemble the ``rough surface'' that \citet{omb_col90} proposed
might reduce the polarization.  As first suggested by \citet{omb_gne78},
Faraday depolarization might also be significant.  The Faraday rotation angle
at the peak of the locally emitted thermal spectrum is roughly
$0.8\tau_{\rm T}R^{1/2}$, where $\tau_{\rm T}\sim1$
is the Thomson depth of the photosphere and $R$ is the ratio of magnetic
energy density to radiation pressure at the photosphere.  For the conditions
observed in the simulation at times of high energy content, this angle turns
out to be approximately 2 radians.

To summarize, the simulation suggests that the vertical structure of a
disk annulus with comparable gas and radiation pressures is as follows.
MRI turbulence churns away at the midplane where the vast majority
of the accretion power is being dissipated as heat.  Away from the midplane,
but still well inside the photosphere, magnetic energy density dominates
both gas and radiation pressures, and magnetic forces provide the hydrostatic
support of the disk against gravity.  These magnetically supported layers are
Parker unstable, resulting in a very complex, inhomogeneous photosphere higher
up.  Outward vertical heat transport is completely dominated by radiative
diffusion, and not by advection, convection, or Poynting flux.  Because most
of the dissipation is occurring at the midplane, no energetically significant
corona is produced above the photosphere.

The absence of a corona was also seen in the gas-pressure dominated
simulation \citep{omb_hir06}, and is a potential problem for models which
suggest that much of the accretion power can be carried locally out of the
disk by magnetic fields (e.g. \citealt{omb_haa91,omb_sve94}).
It might be that the simulation box is too small
to allow enough magnetic field to get out and dissipate to produce a corona.
(It turns out that the wavelength of the fastest growing Parker mode barely fits
inside the box.)  However, this may not change the fact that most of the
dissipation occurs in the MRI turbulent regions near the midplane.  It is
important to bear in mind, though, that this dissipation is mostly numerical
in nature, being caused by losses of magnetic and kinetic energy at the grid
scale.  Still this may be okay if MRI turbulence always cascades energy
down from large to small length scales.

If the simulation results are correct, then where do the X-ray emitting
regions in AGN come from?  It may be that things change when radiation pressure
dominates gas pressure at the midplane.  Magnetic buoyancy may be enhanced
in radiation pressure dominated environments \citep{omb_sak89}, and perhaps
this would alter the vertical dissipation profile to put more power out in
a corona.  However, this was not observed in the preliminary simulation by
\citet{omb_tur04}.  It could also be that the X-ray source has nothing at
all to do with the local disk, but is instead material originating from the
innermost disk and outflowing in a jet or backflowing over the rest of the
disk, as is seen in global, nonradiative, general relativistic MHD simulations
(e.g. \citealt{omb_haw02}).  Regions of high electric current density are
also seen near and inside the plunging region in such simulations
\citep{omb_hir04}, and the dissipation associated with such high currents
might very well produce an X-ray emitting coronal plasma.

I hope I have given the reader a taste of the progress we have made in
understanding the fundamental physical processes inside accretion disks.
Radiation pressure dominated simulations are now in the works!

\acknowledgements 
I am very grateful to my collaborators (Shane Davis, Shigenobu Hirose, Ivan
Hubeny, Julian Krolik, and Neal Turner) without whom the theoretical results
presented here would not have been possible.  I also wish to thank
Erin Bonning, Xinyu Dai, David Pooley, and Greg Shields for sharing their
results with me prior to publication.
This work was supported in part by the U.S. National Science Foundation under
grant AST-0307657.


\end{document}